\documentclass[conference]{IEEEtran}
\IEEEoverridecommandlockouts
\usepackage[english]{babel}
\usepackage[utf8]{inputenc}
\usepackage[T1]{fontenc}
\usepackage{cite}
\usepackage{amsmath,amssymb,amsfonts}
\usepackage{algorithmic}
\usepackage{graphicx}
\usepackage{textcomp}
\usepackage{xcolor}
\usepackage{booktabs}
\usepackage{makecell}
\usepackage{microtype}
\usepackage{graphicx}
\usepackage{amsmath}
\usepackage{xcolor}
\usepackage{hyperref}
\usepackage{siunitx}
\usepackage{mwe}
\usepackage[font=small,labelfont=bf]{caption}
\usepackage{tikz}
\usetikzlibrary{matrix,fit,positioning,shapes.misc,calc}
\usepackage{orcidlink}
\usepackage{fancyhdr}

\begin{document}
\title{Toward GPU-Resident Climate Models: A Feasibility Study on Lossy Compression for the Spherical Harmonic Transform's Communication Bottleneck
  \thanks{This work was partially funded under the NRRP, Mission 4 Component 2 Investment 1.4, by the European Union -- NextGenerationEU (proj. no. CN 00000013, CUP no. E63C22000970007). }
}
\author{\IEEEauthorblockN{Lorenzo Breschi\,\orcidlink{0009-0004-9616-6810}}
  \IEEEauthorblockA{\textit{University of Trento} \\
    Trento, Italy \\
    lorenzo.breschi@unitn.it}
  \and
  \IEEEauthorblockN{Flavio Vella\,\orcidlink{0000-0002-5676-9228}}
  \IEEEauthorblockA{\textit{University of Trento} \\
    Trento, Italy \\
    flavio.vella@unitn.it }
}
\maketitle
\thispagestyle{fancy}
\lhead{}
\rhead{}
\chead{}
\lfoot{\footnotesize{
SC26 Workshops, November 15-20, 2026, Chicago, Illinois, USA
\newline 979-8-3195-1221-5/26/\$31.00 \copyright 2026 IEEE}}
\rfoot{}
\cfoot{}
\renewcommand{\headrulewidth}{0pt}
\renewcommand{\footrulewidth}{0pt}

\begin{abstract}
  Operational pseudospectral atmospheric models such as the ECMWF Integrated Forecasting System (IFS) run today almost exclusively on CPUs; GPU ports are under active development but not yet used in production. These models rely on the Spherical Harmonic Transform (SHT). Each time-step requires forward and inverse SHTs, and both passes depend on global pencil transposition that redistribute multi-dimensional arrays across compute nodes. At large node counts these global collectives dominate wall-clock time. We investigate GPU-resident lossy compression, using representative fields from the DYAMOND high-resolution operational dataset as input, and combining measured GPU compression throughput with SimGrid network simulation, we show that ZFP at 16 bits per value (rate-16), about the same storage budget as float16, matches the communication-time reduction of float16 truncation while delivering approximately $\mathbf{1600\times}$ lower mean relative error. ZFP at 8 bits per value (rate-8) achieves approximately $\mathbf{1.93\times}$  the speedup of float16 while retaining $\mathbf{4\times}$ lower mean relative error.
      
\end{abstract}

\begin{IEEEkeywords}
  spherical harmonic transform, lossy compression, GPU computing, numerical weather prediction, pseudospectral models, all-to-all communication, mixed precision, high-performance computing
\end{IEEEkeywords}

\section{Introduction}
\label{sec:intro}
Numerical weather prediction (NWP) and global climate modeling
perform some of the largest and most time-critical simulations in
computational science.  One of the dominant operational approach to global
forecasting is the pseudospectral discretization, exemplified by the
ECMWF Integrated Forecasting System (IFS), which today remains a
CPU-only production system. GPU ports of IFS and of the wider ECMWF
modeling infrastructure are under active development, but as of the
most recent ECMWF report they are not yet used operationally
\cite{ecmwf_2026__ECMWFNewsletter186Winter}.  We focus on
pseudospectral models specifically because they are
communication-bound relative to the grid-point (nearest-neighbor)
dynamical cores used by other models, such as ICON
\cite{klocke_2025_ICONat1kmGordonBell_ComputingFullEarthSystem}.
The Spherical Harmonic Transform
(SHT) is at the heart of a pseudospectral model and it requires two global
all-to-all-like data redistributions per time-step (Section
\ref{sec:pencil}).  This makes inter-node communication, rather than
local compute, the eventual bottleneck to GPU-accelerated pseudospectral
forecasting and it is precisely this bottleneck that motivates the
present study.
This pseudospectral core, common to IFS and related models, employs
the Spherical Harmonic Transform (SHT) to compute spatial derivatives
and to move between spectral coefficients and physical-space grid
values.  Because
the SHT is based on a Gauss--Legendre quadrature in latitude and a
Fourier decomposition in longitude, each time-step requires a forward
and an inverse SHT.  On a distributed-memory machine these transforms
involve two global data redistribution, so-called \emph{pencil
  transpositions}, that rearrange multi-dimensional arrays via
all-to-all collectives: one by rows (from $z$-pencils to
$\lambda$-pencils before the longitude FFT) and one by columns (from
$\lambda$-pencils to $\theta$-pencils before the Legendre transform)(see Section \ref{sec:pencil} for details).
These pencil transpositions scale poorly on modern clusters.  While
GPU floating-point throughput has grown by orders of magnitude over the
past decade, inter-node network bandwidth and latency
have not kept pace \cite{desensi_2024_ExploringGPU-to-GPUCommunication_ExploringGPUtoGPUCommunication2024,unat_2024__LandscapeGPUCentricCommunication}.
As the number of nodes grows, the time spent in global all-to-all
communication becomes the dominant contributor to wall-clock time,
limiting strong-scaling efficiency and eroding the benefit of faster
local computation.

A straightforward mitigation is to halve message sizes by casting
exchanged buffers to half-precision float (float16) before
transmission and restoring float32 on the receiver.  This is
essentially a form of mixed-precision communication and has attracted
growing interest in the NWP community \cite{vana_2017_SinglePrecisioninWeatherForecastingModels_SinglePrecisionWeatherForecasting}.  Its appeal is simplicity: no external library,
negligible cast overhead, and a guaranteed 2$\times$ reduction in
message size.  However, the resulting truncation error is
uncontrolled, cannot be bounded without knowledge of the data, and may
differ dramatically across atmospheric variables and vertical levels.
Float16 truncation can introduce errors that accumulate across time-steps and
degrade forecast accuracy.
We propose \emph{online GPU-resident lossy compression} as a
principled alternative.  In our approach, each MPI rank compresses its
outgoing communication buffer on the GPU immediately before the
all-to-all, transmits the compressed byte stream (of fixed,
predictable size), and decompresses on the receiving side before any
local computation.  By performing compression and decompression on the
GPU, we avoid costly host--device memory transfers.  By using ZFP in
fixed-rate mode, we obtain deterministic output
sizes compatible with standard MPI semantics while exploiting spatial
correlations in the atmospheric data to deliver substantially tighter
error bounds than float16 for the same storage budget.

This paper makes the following contributions:
\begin{enumerate}
  \item \textbf{Characterization of NWP data compressibility:}  We
        analyze representative DYAMOND \cite{stevens_2019_DYAMOND_DYAMONDDYnamicsAtmosphericgeneral}
        float32 fields and characterize the relative
        error achieved by ZFP fixed-rate compression at multiple rates,
        comparing against float16 truncation.
  \item \textbf{Feasibility study of compressed pencil communication:}
        To our knowledge, this is the first study to estimate the
        benefit of GPU-resident compressed communication for a
        pseudospectral climate/NWP code, which today runs on CPU only.
        We implement a realistic benchmark of the SHT's first
        all-to-all transposition, measuring GPU compression and
        decompression throughput on real hardware and modeling
        end-to-end communication time with SimGrid \cite{casanova_2025_SimGrid_LoweringEntryBarriersdeveloping} on a Dragonfly topology at node counts from 4 to 121,
        representative of current operational NWP deployments.
  \item \textbf{Quantitative comparison with float16:}  We show that
        ZFP rate-16 matches the communication-time reduction of float16
        while delivering a mean relative error of $2.5 \times 10^{-7}$
        versus $4 \times 10^{-4}$ for float16 (a $1600\times$
        improvement) and that ZFP rate-8 achieves $1.93\times$
        the speedup of float16 at a mean relative error of $10^{-4}$ (a
        $4\times$ improvement).
          
  \item \textbf{Real-hardware validation:}  We corroborate the
        simulated communication-time trend with measurements on the CPU
        partition of the CRESCO8 cluster, grounding the study in a
        currently-deployed, production-representative system
        (Sections \ref{sec:cresco8} and \ref{sec:results_cresco8}).
\end{enumerate}
The remainder of the paper is organized as follows.
Section \ref{sec:background} provides background on the Spherical Harmonic Transform, pencil transpositions, and ZFP compression.  Section \ref{sec:related} reviews related work on SHT libraries, distributed FFTs, mixed precision in NWP, and scientific data compression.  Section \ref{sec:method} describes our experimental methodology, including data selection, benchmark design, and performance modeling.  Section \ref{sec:results} presents our main findings on compressibility and communication time reduction. Some future directions are discussed in Section \ref{sec:future_directions}. Finally, Section \ref{sec:conclusion} summarizes our contributions.
\section{Background}
\label{sec:background}
\subsection{Spherical Harmonic Transform}
\label{sec:sht}
The Spherical Harmonic Transform decomposes a function defined on the
sphere into a linear combination of spherical harmonics
$Y_\ell^m(\theta, \lambda)$, where $\theta$ is the co-latitude, $\lambda$
is the longitude, and $\ell, m$ are the degree and order.  For
numerical weather prediction, the SHT provides a spectrally accurate
way to evaluate horizontal derivatives and to apply spectral filters or
semi-implicit solvers \cite{coiffier_2011_FundamentalsofNWPBook_FundamentalsNumericalWeatherPrediction}
Computationally, the SHT decomposes into a longitude transform (a
standard FFT) and a latitude transform based on the associated Legendre
polynomials $P_\ell^m(\cos\theta)$.  The Gauss--Legendre quadrature
scheme we consider uses Gauss--Legendre latitudes to achieve exact
quadrature for polynomials up to the truncation wavenumber; this is the
scheme used in operational models such as the ECMWF IFS, as distinct
from equispaced schemes that are more common in data-analysis
applications.  In practice, atmospheric state variables are held on a
three-dimensional grid with $N_\lambda$ longitude points, $N_\theta$
Gauss--Legendre latitude points, and $N_z$ vertical levels.
\subsection{Pencil Transpositions}
\label{sec:pencil}
On a distributed-memory system, the 3-D data array is decomposed
across MPI ranks.  In the \emph{pencil decomposition}, the
domain is partitioned along two of its three dimensions, so that each
rank holds a pencil aligned with the third, as illustrated in
Figure \ref{fig:pencil_comm}.  Because the FFT (in
longitude) and the Legendre transform (in latitude) must each operate
along their respective dimension without distributed data, the
transform requires two global redistributions:
\begin{figure}[t]
  \includegraphics[width=\columnwidth]{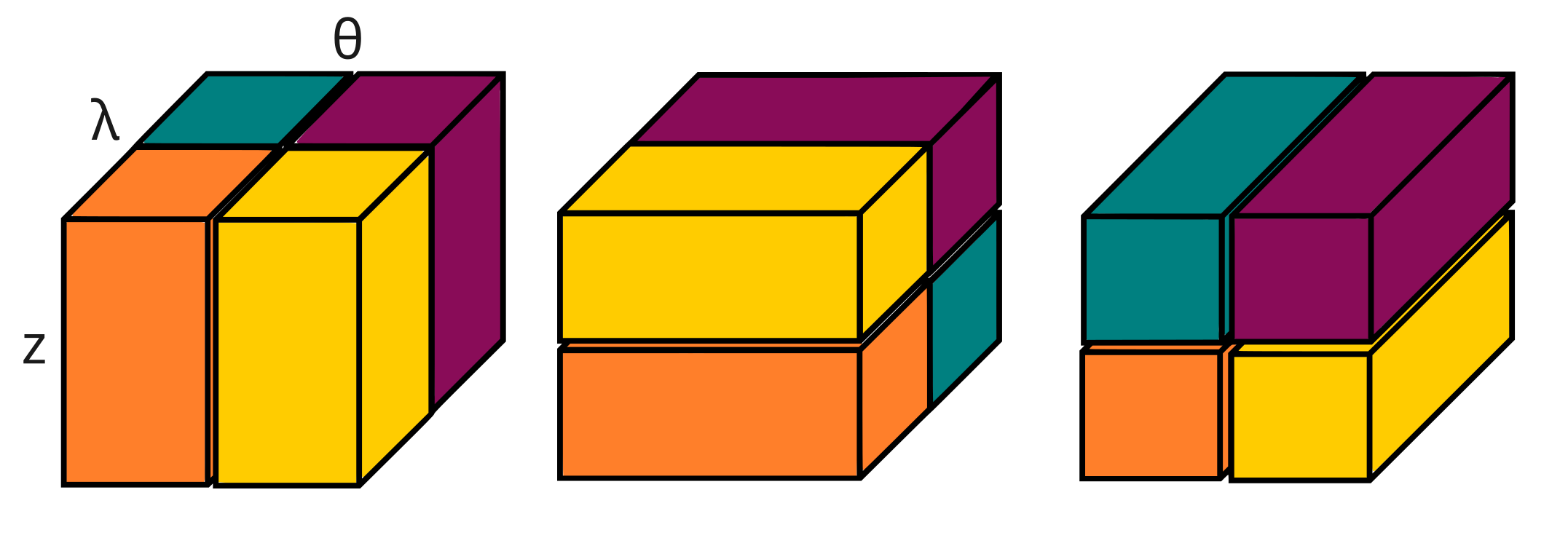}
  \caption{Pencil decomposition of a 3-D data array across MPI ranks.
    Each rank owns a contiguous slab (pencil) along one axis.
    A pencil transposition redistributes ownership among ranks,
    realigning the pencils with a different axis so that the
    next transform (FFT or Legendre) can proceed locally.
    The redistribution is implemented as an all-to-all
    collective  by rows/columns in which every rank simultaneously sends to and
    receives from every other rank.}
  \label{fig:pencil_comm}
\end{figure}

\begin{enumerate}
  \item \textbf{All-to-all by rows} (transpose from $z$-pencils to
        $\lambda$-pencils): each rank sends one horizontal slice to
        every other rank, enabling the longitude FFT.
  \item \textbf{All-to-all by columns} (transpose from $\lambda$-pencils
        to $\theta$-pencils): a second collective enables the latitude
        Legendre transform.
\end{enumerate}
Both transpositions are local (row-wise or column-wise) all-to-all calls
in which every rank is both a sender and a receiver.
In this paper we study the first transposition (rows),
noting that the second is structurally analogous.
The communication bottleneck arises because collective bandwidth
scales with the number of communicating ranks while per-link network
bandwidth does not.  On clusters where GPU compute throughput is
plentiful but inter-node bandwidth is limited, a gap that continues to
widen, the all-to-all dominates
wall time at moderate-to-large node counts \cite{desensi_2024_ExploringGPU-to-GPUCommunication_ExploringGPUtoGPUCommunication2024,unat_2024__LandscapeGPUCentricCommunication}.
\subsection{ZFP Fixed-Rate Compression}
\label{sec:zfp}
ZFP \cite{lindstrom_2014_ZFP_FixedRateCompressedFloatingPointArrays,diffenderfer_2019_ZFPanalysis_ErrorAnalysisZFPCompression} is an open-source compressor designed
for multi-dimensional floating-point arrays that exhibit spatial
correlation, as is typical of regularly sampled continuous fields.
ZFP partitions the input array into non-overlapping $4^d$ blocks (where
$d$ is the array dimensionality) and compresses each block independently
by applying a reversible lifting transform followed by a precision
reduction.
ZFP offers three compression modes.  The \emph{accuracy} mode
guarantees that every reconstructed value is within a user-specified
absolute error tolerance, but produces variable-size output that is
incompatible with standard MPI pre-allocation and is not well-suited to
GPU parallelism.  The \emph{precision} mode targets a fixed number of
uncompressed bits per value.  The \emph{fixed-rate} mode assigns
exactly $r$ compressed bits per value, producing output of
deterministic size and enabling trivially parallel compression of
independent blocks on a GPU.
For our initial investigation, we chose to use ZFP in fixed-rate mode.
Other compression libraries for scientific data offer different tradeoffs in their accuracy models, performance profiles, and GPU support (see Section \ref{subsec:compression_scientific_data}). Those will be explored in a future work.

\section{Related Work}
\label{sec:related}
\subsection{SHT Libraries for Distributed Systems}
\label{subsec:sht_libraries}
Several open-source SHT libraries are available, but few support
both MPI distribution and GPU acceleration simultaneously.
SHTns \cite{schaeffer_2012_SHTns_EfficientSphericalHarmonicTransforms} and DUCC \cite{reinecke_2020_DUCC_DUCCDistinctlyUsefulCode} are highly optimized for shared-memory
machines (CPU+GPU) but do not support MPI.  HEALPix \cite{gorski_2005_HEALPix_HEALPixFrameworkHighResolutionDiscretization}, CHarm \cite{bucha_2026_CHarm_CHarmPythonLibraryspherical}
and related astronomical codes operate on a different pixelization
scheme and typically analyze a single scalar field, making them
unsuitable for NWP with hundreds of vertical levels. ISPACK \cite{ishioka_2018_ISPACK_NewRecurrenceFormulaEfficient} and Libsharp \cite{reinecke_2013_Libsharp_LibsharpSphericalHarmonictransforms} are consolidated libraries that have MPI support but lack GPU support.
The ECMWF ecTrans library \cite{ecmwf_2024_IFSDocumentationCY49R1-PartVI_IFSDocumentationCY49R1Part} is, to our knowledge, the only library
suitable for large-scale operational NWP: it supports MPI distribution,
operates on Gauss--Legendre grids, and has been partially ported to
GPU.  However, as of the time of writing, operational
ECMWF forecasts are still produced on CPUs, and the GPU port has not
yet been fully optimized for production use \cite{ecmwf_2026__ECMWFNewsletter186Winter}
\subsection{Distributed FFT and Pencil Communication}
\label{subsec:fft_libraries}
The pencil transposition pattern is essentially identical to that used
in distributed three-dimensional FFTs.  Libraries such as
2DECOMP\&FFT\cite{li_2010_2decomp-fft_2decompFftaHighlyscalable,rolfo_2023_2decomp-fft2_2DECOMPFFTLibraryUpdatenew}, P3DFFT\cite{pekurovsky_2012_P3DFFT_P3DFFTFrameworkParallelcomputations},
and heFFTe \cite{ayala_2020_heFFTe_HeFFTeHighlyEfficientFFT}
implement pencil transpositions for FFTs, and their communication
structure and scaling behavior are directly relevant to the SHT
setting.  None of these libraries currently support
variable-length compressed messages or expose a compression hook in
their communication pipeline.
2DECOMP\&FFT\cite{li_2010_2decomp-fft_2decompFftaHighlyscalable,rolfo_2023_2decomp-fft2_2DECOMPFFTLibraryUpdatenew} exposes an API for general pencil transposition and
cuDecomp\cite{romero_2022_CuDecomp_DistributedmemorySimulationsTurbulentflows} is a library focused on GPU-accelerated pencil transpositions with an automatic choice of communication strategy and backend.
A closely related contribution to our work is the MFFT library\cite{zhao_2023_MFFT_MFFTGPUAcceleratedHighly},
which applies lossy compression to the inter-process communication
buffers of a distributed FFT.  MFFT reports performance improvements
over 2DECOMP\&FFT (however MFFT uses GPU and 2DECOMP\&FFT uses CPU) and heFFTe (GPU-enabled, but MFFT and heFFTe have a different data layout and heFFTe needs three communication steps as opposed to two), using compression to reduce message sizes. However, the comparison is somewhat unbalanced as MFFT uses GPUs for the computation while 2DECOMP\&FFT and heFFTe use CPUs. As for heFFTe, the comparison is not entirely fair as heFFTe uses a different data layout (blocks as opposed to pencils) and needs three communication steps as opposed to two, and communication is known to be the bottleneck for heFFTe \cite{ayala_2020_heFFTe_HeFFTeHighlyEfficientFFT}.
The key difference between this work and MFFT is that MFFT focuses on FFT and uses a custom lossy compression scheme, while we focus on SHT and uses ZFP. Our work is preliminary and a comparison with MFFT is left for future work. 
Cayrols et al. \cite{cayrols_2022_TruncatedheFFTe_LossyAlltoallExchangeaccelerating} have studied the speedup of float16 truncation during communication for FFT, while they achieve a significant speedup, the approach is naive and not tunable, as the error cannot be controlled. Furthermore, the accuracy gain of using float32 computation with float16 communication over using both computation and communication in float16 may not justify this approach.

\subsection{Mixed Precision in NWP}
\label{subsec:mixed_precision}
The use of reduced floating-point precision in atmospheric models has
received considerable attention.  The ECMWF IFS now runs
parts of its computation in single precision \cite{vana_2017_SinglePrecisioninWeatherForecastingModels_SinglePrecisionWeatherForecasting}.  Our work is complementary: rather than reducing the
precision of arithmetic, we reduce the precision of
\emph{communication} buffers, using a compressor that adapts its
bit-allocation to local field structure.
\subsection{Compression of NWP Data}
\label{subsec:compression_nwp_data}
Various studies have investigated the use of lossy compression for NWP data, but they have focused on \emph{archival} compression of large datasets, not on online compression of communication buffers.
Tint\'o Prims et al. \cite{tintoprims_2024__EffectLossyCompressionnumerical} study the effect of lossy compression of whole datasets for analysis purposes with a custom compressor. Huang et al. \cite{huang_2023__CompressingMultidimensionalWeatherclimate} study the feasibility of custom ML compressors
for archival purposes.
Those works
exploit temporal correlations across many time-steps and long records
and are unsuitable for online per-message compression.  Our setting is
fundamentally different: we compress a small spatial patch of a single
time-step, in real time, on a GPU, with no access to past or future
fields.
Kl\"ower et al. \cite{klower_2021__CompressingAtmosphericDataits} do a study on the effect of lossy compression on the CAMS dataset (float64) using ZFP as the present work. However, their work focuses on the effect of lossy compression on accuracy and not on the communication time, which is the focus of our work. They study the compression of a single time-step, without knowledge of past and future data, but they still compress data for the whole 3D atmospheric fields. In our work we are bound to compression on data resident on a single GPU, and so the achievable compression ratio is limited. Their work however is complementary to ours, and we plan to extend our work by using their analysis of the effect of lossy compression on the forecast accuracy to inform the choice of compression parameters for the best tradeoff between communication time and forecast accuracy.

\subsection{Compression of Scientific Floating-Point Data}
\label{subsec:compression_scientific_data}
Many compressors exist for scientific data, including ZFP \cite{lindstrom_2014_ZFP_FixedRateCompressedFloatingPointArrays}, SZ and its variants \cite{liang_2023_SZ3_SZ3ModularFrameworkComposing,tian_2020_cuSZ_CuSZEfficientGPUBasedErrorBounded},
and MGARD\cite{gong_2023_MGARD_MGARDMultigridFrameworkhighperformance}. See also Di et al. \cite{di_2025__SurveyErrorBoundedLossyCompression} for a survey.
Compressor in their accuracy models, performance profiles, and GPU support.  For our
feasibility study the choice of compressor is secondary; ZFP was
selected because its CUDA backend (cuZFP) is mature, its fixed-rate
mode produces outputs of predictable size, it has been extensively
validated in HPC settings \cite{diffenderfer_2019_ZFPanalysis_ErrorAnalysisZFPCompression}, and has already been used for compression of atmospheric data \cite{klower_2021__CompressingAtmosphericDataits}.
The results we report are therefore a conservative lower bound on what a
compressor better adapted to atmospheric field structure could achieve.
\subsection{Compressed Communication Middleware}
\label{subsec:compressed_middleware}
Recent works on compression-aware MPI and network middleware \cite{zhou_2022__AcceleratingMPIAlltoAllCommunication,zhou_2021__DesigningHighPerformanceMPILibraries,huang_2024__OptimizedErrorcontrolledMPICollective,ma_2026_UCCL-Zip_UCCLZipLosslessCompressionSupercharged} apply generic compression algorithms at the communication layer.  Such approaches do not necessitate to modify application code, but they are not suitable for our use case.  These compressors are typically
data-agnostic and operate on flattened one-dimensional byte buffers,
which prevents them from exploiting the multi-dimensional block
structure on which accuracy guarantees of scientific compression libraries depend.
Because MPI primitives do not expose multi-dimensional array
metadata to the communication layer, injecting a structure-aware
compressor requires either application-level wrapping (our approach)
or substantial MPI implementation changes.

\section{Methodology}
\label{sec:method}
\subsection{Experimental Data}
\label{subsec:data}
We use output from the DYAMOND \cite{stevens_2019_DYAMOND_DYAMONDDYnamicsAtmosphericgeneral} high-resolution
global storm-resolving simulation at $2560 \times 1281$ grid points,
approximately twice the horizontal resolution of current ECMWF
operational forecasts.  The dataset is distributed in 32-bit
single-precision floating point, making it representative of the
precision used in modern NWP systems (operational models are
increasingly moving from float64 to float32 \cite{vana_2017_SinglePrecisioninWeatherForecastingModels_SinglePrecisionWeatherForecasting}).
Operational NWP fields fall into two distinct categories.
Dynamical-core variables such as temperature and ozone are stored and
communicated as native float32.  Many tracer and diagnostic variables,
however, are stored with an integer quantization scheme (an integer
array paired with a scale factor and offset), which reduces their
dynamic range and makes them unsuitable test cases for a floating-point
compressor: the piecewise-constant quantization creates sharp step
discontinuities that are pathological for block-transform compressors
such as ZFP.  We therefore focus on temperature ($T$) and ozone
($O_3$), which are representative of the float32 dynamical-core
variables that dominate communication volume and forecast sensitivity.
We focus on compressing floating point variables as we are interested in the performance of compressing data for numerical prediction (online), as opposed to compressing data for archival purposes (offline).
\subsection{Compression Configuration}
\label{sec:compression}
We use ZFP version 1.0.1 in fixed-rate mode.
Fixed-rate mode was selected for an initial performance investigation because it is the only mode with GPU support and because it produces messages of deterministic size, which is compatible with standard MPI semantics and easier to integrate into existing SHT libraries.  Future works may explore accuracy-mode compression, which offers tighter error control at the cost of variable message sizes.
We evaluate two rates.  \textbf{Rate-16} (16 bits per value) produces
messages of the same size as float16 truncation, enabling a direct
comparison of accuracy at equal communication cost.  \textbf{Rate-8}
(8 bits per value) halves message size again, at the cost of higher
compression error.
\subsection{GPU Compression Throughput}
\label{sec:gpu_throughput}
Compression and decompression were performed on a NVIDIA A100 SXM4 GPU (\SI{19.5}{TFlop/s} of theoretical float32 performance) with \SI{40}{GB} of memory, using the cuZFP CUDA backend.
We measured wall-clock throughput by using the cuZFP CUDA backend.  We measured wall-clock throughput by
compressing and decompressing the message buffers for representative fields $T$ and
$O_3$.  Measurements were taken after a warm-up phase to
exclude GPU initialization overhead.

CPU compression throughput was also measured on a AMD EPYC 7742 64-Core Processor (
\SI{2.5}{GHz} of clock frequency) for
comparison. Compression is compute-bound and multi-core CPU does not provide the necessary parallelism for efficient compression.  We present this
result quantitatively in Section \ref{sec:results_cpu} to motivate the
GPU-only approach.

\subsection{Network Simulation with SimGrid}
\label{sec:simgrid}

We use state-of-the-art SimGrid simulator \cite{casanova_2025_SimGrid_LoweringEntryBarriersdeveloping} with its SMPI module to model all-to-all communication time
at scale.  Its key advantage for our purposes is that it
eliminates measurement noise arising from production network congestion,
enabling clean comparisons between communication strategies across a
wide range of node counts without requiring access to a large cluster.
We modeled a Dragonfly topology with the following parameters: 27 groups with 4 group-group links,
16 chassis per group with 3 chassis-to-chassis links, 4 routers per
chassis with 1 router-to-router link, and 1 nodes per router, yielding
6480 nodes in total. Link bandwidth was set to \SI{200}{Gbps} with a
latency of \SI{1}{\micro s} per hop.  This topology is similar in
scale and configuration to the JUPITER system. Node counts ranged from 4 to
121 nodes (as listed in Table \ref{table:message_size}), a range
directly relevant to current operational NWP deployments.  We chose
this range rather than a larger future-scale simulation to demonstrate
that the compression approach is beneficial for present-day routine weather forecasts, not only
at hypothetical exascale node counts.
\begin{figure}[hb]
  \begin{tabular}{r|rrrr}
    \toprule
    \makecell{$N$                          \\(nodes)} &
    \makecell{Float32                      \\(MB)} &
    \makecell{Float16                      \\(MB)} &
    \makecell{Rate-16                      \\(MB)} &
    \makecell{Rate-8                       \\(MB)} \\
    \midrule
    4   & 212.50 & 106.25 & 106.25 & 53.13 \\
    9   & 62.38  & 31.19  & 33.54  & 16.77 \\
    16  & 26.56  & 13.28  & 14.06  & 7.03  \\
    25  & 13.50  & 6.75   & 7.00   & 3.50  \\
    36  & 7.62   & 3.81   & 4.23   & 2.12  \\
    49  & 4.81   & 2.41   & 2.58   & 1.29  \\
    64  & 3.32   & 1.66   & 1.95   & 0.98  \\
    81  & 2.31   & 1.15   & 1.25   & 0.62  \\
    100 & 1.62   & 0.81   & 1.00   & 0.50  \\
    121 & 1.23   & 0.62   & 0.62   & 0.31  \\
    \bottomrule
  \end{tabular}
  \captionof{table}{MPI message sizes per rank-pair (in MB) for the all-to-all-by-rows
    transposition as a function of node count $N$.  Sizes are
    computed for the DYAMOND $2560 \times 1281$ grid.  Float32 is the
    uncompressed baseline; float16 and ZFP rate-16 are
    similar in size by construction (both store 16 bits per
    value); ZFP rate-8 approximately halves that again.}
  \label{table:message_size}
\end{figure}

The simulation proceeds as follows.  For each node count $N$ and each
compression strategy, SimGrid simulates the communication
collective using the message sizes from Table \ref{table:message_size}.
The compression and decompression overheads, are added as a
compute-side delay on the sender (compression) and receiver
(decompression) before and after the collective, respectively.  Float16
truncation is modeled with negligible cast overhead (the cast from
float32 to float16 is a GPU register operation).  Total time is
reported as $t_\mathrm{compress} + t_\mathrm{communication} +
  t_\mathrm{decompress}$, where $t_\mathrm{communication}$ is the simulated
communication time and the remaining terms are measured.
This serial model is a conservative upper bound, as compression and communication could in principle be pipelined per block.
\subsection{CRESCO8 Real-Hardware Measurements}
\label{sec:cresco8}
To complement the SimGrid simulation, we measured communication time
directly on the CPU partition of the CRESCO8 cluster (ENEA, Portici,
Italy).  CRESCO8 CPU partition comprises 760 compute nodes (dual-socket, 64-core Intel Xeon Platinum 8592+,
1.9--3.8\,GHz, 512\,GB RAM)
Nodes are connected through a 200\,Gb/s Mellanox NDR InfiniBand fabric
in a Dragonfly topology, the same topology class used in the SimGrid
model.
We ran the row-wise all-to-all transposition on CRESCO8's CPU nodes
and measured communication time for the four strategies (float32,
float16, ZFP rate-16, ZFP rate-8) at world sizes of 4, 16, and 64.
Operational NWP codes run on CPU only, so compression and
decompression segments are not measured natively; they reuse the
NVIDIA A100 throughput of Section \ref{sec:gpu_throughput}, while only
the communication segments were measured on CRESCO8.
\subsection{Error Metric}
\label{sec:error_metric}
As a proxy for the error on the downstream application, we report the \emph{relative
  error}:
\[
  \epsilon =
  \frac{\|x - \hat{x}\|}{\|x\|},
\]
where $x$ are the original values and $\hat{x}$ are the reconstructed values.
For float16 truncation the analogous maximum relative error is computed by casting the float32 values to
float16 and back and applying the same formula.

\section{Results}
\label{sec:results}
\subsection{SimGrid Simulation of Communication Time}
\label{sec:results_time}
\begin{figure}[h]
  \includegraphics[width=\columnwidth]{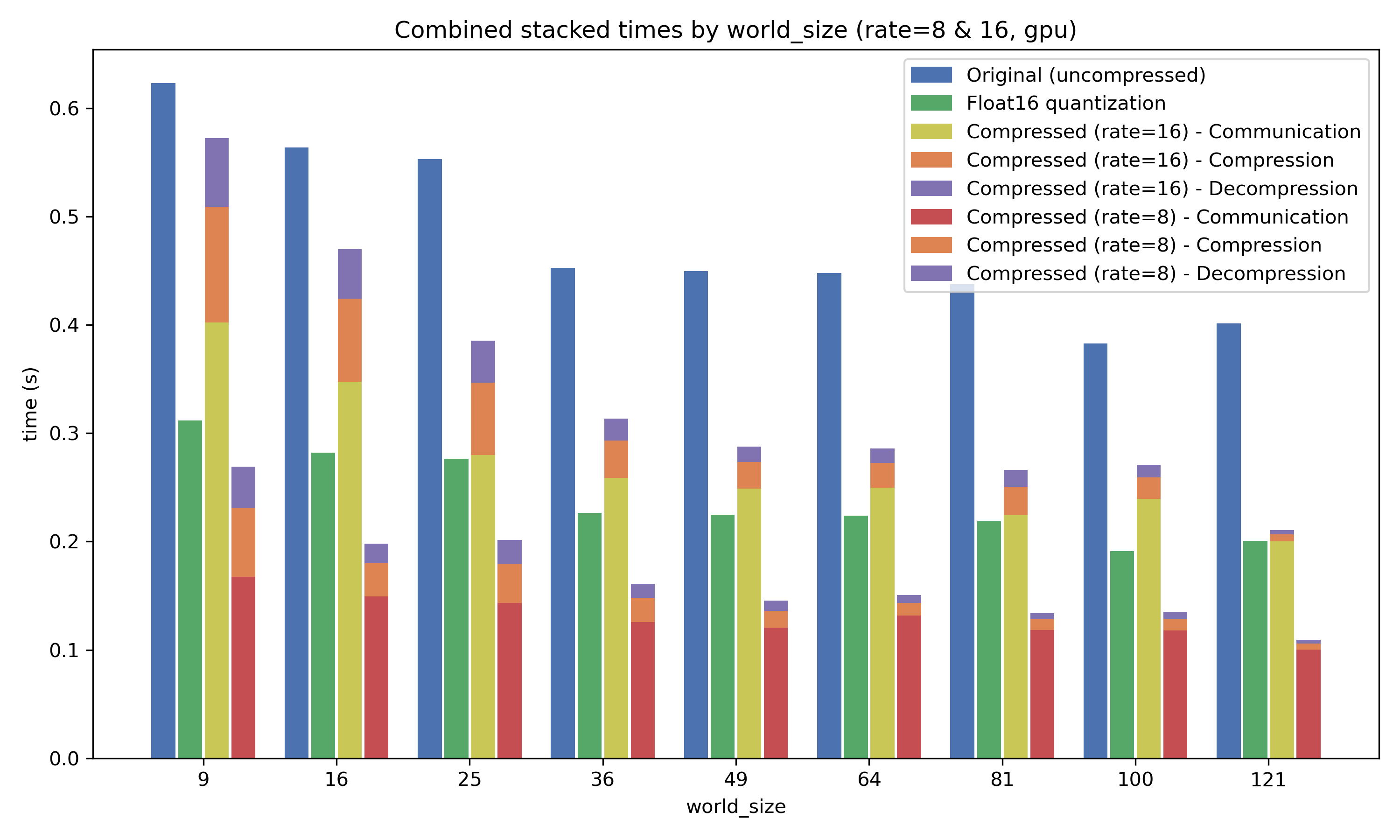}
  \caption{Simulated end-to-end time (compression + communication +
    decompression) for the four strategies, as a function of
    node count $N$ from 4 to 121, using GPU compression and
    decompression throughput measured on the NVIDIA A100
    (Section \ref{sec:gpu_throughput}).  The compression and
    decompression overhead is a small and shrinking fraction of
    the total as $N$ grows, becoming almost negligible at the
    largest node counts.}
  \label{fig:stacked_gpu}
\end{figure}

Figure \ref{fig:stacked_gpu} shows stacked bar plots of total time
(compression overhead + communication time + decompression overhead)
as a function of node count $N$, for four strategies: baseline float32,
float16 truncation, ZFP rate-16, and ZFP rate-8.  The following
observations summarize the results.
\textbf{Rate-16 vs.\ float16.}  At small node counts ($N \leq 25$),
GPU compression overhead is non-negligible relative to the communication
time, and rate-16 is slightly slower than float16.  At
$N \approx 36$--$100$ nodes, the compression overhead
becomes small relative to the reduced communication time and rate-16
matches float16.  At large node counts ($N = 121$), the message size of float16 and rate-16 is comparable and compression overhead becomes negligible, so the speedup of both strategies is similar.
This matched speedup comes at a much lower error: as Section
\ref{sec:results_error} shows, rate-16's mean relative error is
$1600\times$ lower than float16's (Figure \ref{fig:err_rate16}).
\textbf{Rate-8.}  Rate-8 produces messages about half the size of rate-16,
and at large node counts ($N \geq 100$) it achieves approximately $1.93\times$ the
speedup of float16 over the float32 baseline.  At all node counts, rate-8 is faster than float16.
As Section \ref{sec:results_error} shows, rate-8 also retains a
$4\times$ lower mean relative error than float16 (Figure
\ref{fig:err_rate8}).

\subsection{CRESCO8 Communication Measurements}
\label{sec:results_cresco8}
\begin{figure}[h]
  \includegraphics[width=\columnwidth]{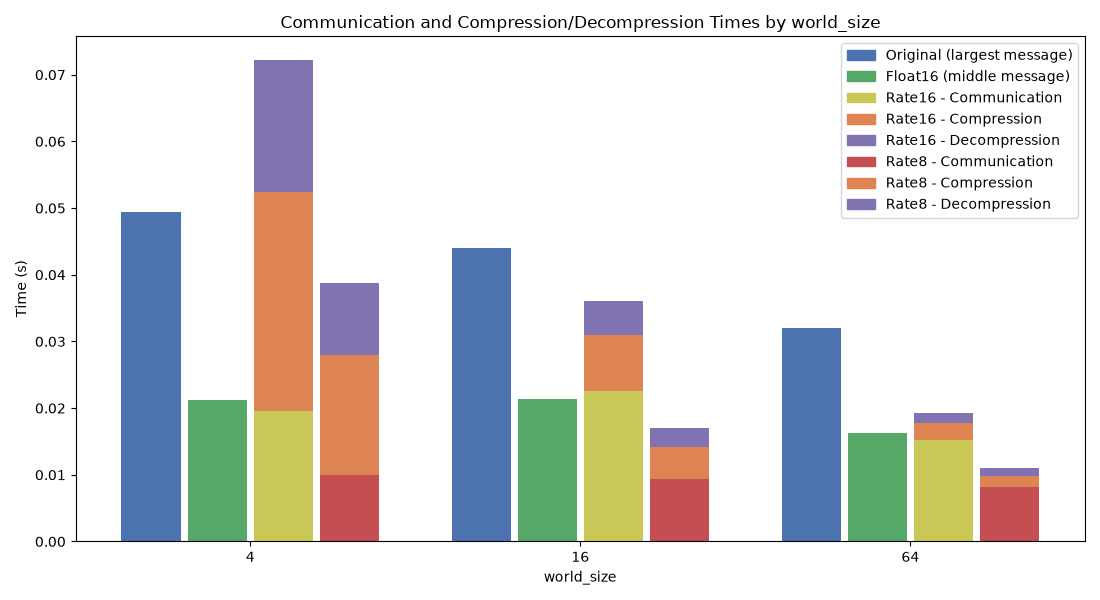}
  \caption{Measured communication time on the CRESCO8 CPU partition
    for the four strategies, as a function of MPI world size (4, 16,
    64).  Compression and decompression segments overlay the NVIDIA
    A100 throughput of Section \ref{sec:gpu_throughput}; only the
    communication segments were measured on CRESCO8 itself.}
  \label{fig:cresco_stacked}
\end{figure}

Figure \ref{fig:cresco_stacked} shows stacked bar plots of total time
(compression overhead + communication time + decompression overhead)
on the CRESCO8 CPU partition (Section \ref{sec:cresco8}) for the four
strategies, as a function of MPI world size.  As in the simulated
results of Figure \ref{fig:stacked_gpu}, the compression/decompression
overhead shrinks relative to communication as scale grows.  The real
measurements reach world size 64.  The following observations
summarize the results.
\textbf{Rate-16 vs.\ float16.}  At $N = 4$, compression overhead
dominates the total time, though this world size is too small to be
representative of realistic deployments.  At $N = 16$, the compressed
message is faster than the uncompressed baseline but not yet
competitive with float16.  At $N = 64$, rate-16 is slightly slower
than float16; however, as Section \ref{sec:results_error} shows,
rate-16's mean relative error is $1600\times$ lower than float16's
(Figure \ref{fig:err_rate16}).
\textbf{Rate-8.}  At $N = 4$, compression overhead again dominates the
total time, and this world size is not representative of realistic
deployments.  At $N = 16$, rate-8 is faster than float16 while
retaining better accuracy.  At $N = 64$, rate-8 is considerably faster
than float16, and, as Section \ref{sec:results_error} shows, retains a
$4\times$ lower mean relative error (Figure \ref{fig:err_rate8}).
\subsection{Data Compressibility and Compression Error}
\label{sec:results_error}
\begin{figure}[h]
  \includegraphics[width=\linewidth]{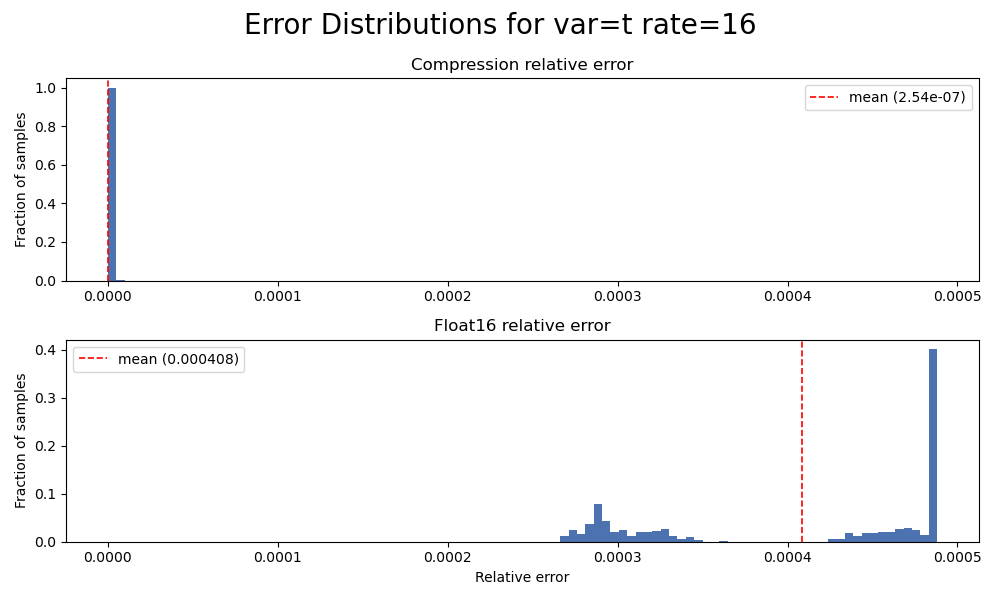}
  \caption{Relative error distributions for ZFP rate-16 and
    float16 truncation applied to the DYAMOND summer temperature field
    (2016-08-01 00Z).  ZFP rate-16 achieves a mean
    relative error of $2.5 \times 10^{-7}$; float16 truncation
    achieves a mean of $4 \times 10^{-4}$. Compression achieves a $1600\times$ lower mean relative error than float16 at the same storage cost. Ozone ($O_3$) shows a very similar relative-error distribution and is omitted for brevity.
  }
  \label{fig:err_rate16}
\end{figure}

\begin{figure}[h]
  \includegraphics[width=\linewidth]{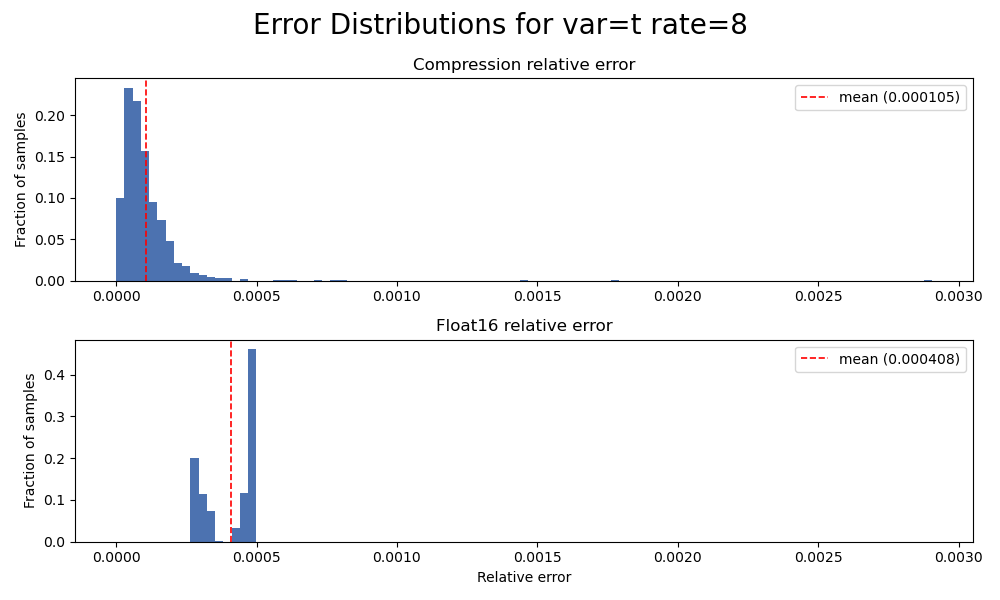}
  \caption{Relative error distributions for ZFP rate-8 and
    float16 truncation applied to the same DYAMOND temperature
    field.  ZFP rate-8, which uses around 8 bits per value (half
    the storage of float16), achieves a mean relative error of
    $10^{-4}$, better than the float16 mean of $4 \times 10^{-4}$. As with rate-16, ozone ($O_3$) shows a very similar relative-error distribution and is omitted for brevity.}
  \label{fig:err_rate8}
\end{figure}

Table \ref{table:message_size}
shows how individual message sizes decrease as node count grows.
Message sizes have been calculated on the DYAMOND data for the temperature field.
Figures \ref{fig:err_rate16} and \ref{fig:err_rate8} show the
distributions of relative error for temperature ($T$) over
all $4^3$ ZFP blocks in the DYAMOND field, comparing ZFP compression
against float16 truncation at rate-16 and rate-8 respectively.
Figure \ref{fig:err_rate16} shows a clear difference at similar
storage cost.  ZFP rate-16 achieves a mean relative error of
$2.5 \times 10^{-7}$, compared to $4 \times 10^{-4}$ for float16
truncation, a factor of $1600\times$ lower.
Figure \ref{fig:err_rate8} shows that ZFP rate-8, using half the
storage of float16, achieves a mean relative error of
$10^{-4}$, within a factor of 4 of float16's mean relative error, while approximately halving the message size again.
This error analysis suggests that for a typical atmospheric field, ZFP's block-transform strategy achieves an acceptable error with a better accuracy--performance trade-off than float16. This may not be the case for extreme events, but this has to be further investigated in future work.
We are aware that the error analysis is limited to a couple of atmospheric variables and a single snapshot. We plan to extend our work by using different fields from various weather condition following the analysis of Kl\"ower et al. \cite{klower_2021__CompressingAtmosphericDataits}.

\subsection{CPU Compression: Why It Is Not Viable}
\label{sec:results_cpu}
\begin{figure}[h]
  \includegraphics[width=\columnwidth]{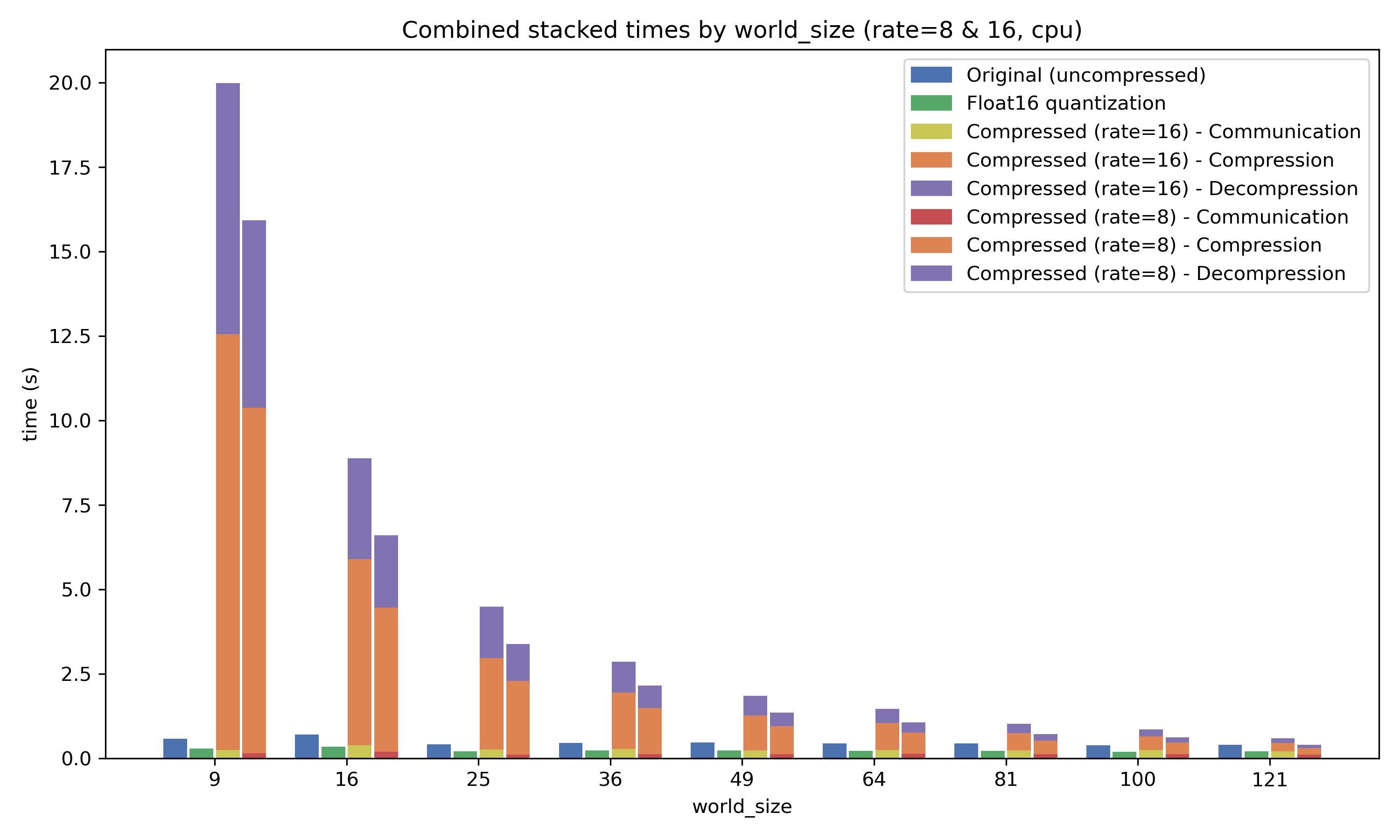}
  \caption{Same breakdown as Figure \ref{fig:stacked_gpu}, using CPU
    compression and decompression throughput (AMD EPYC 7742)
    instead of GPU throughput.  CPU compression time exceeds the
    communication time for most node counts, only becoming
    comparable to it at the largest node counts.}
  \label{fig:stacked_cpu}
\end{figure}

Figure \ref{fig:stacked_cpu} shows a stacked bar chart similar to the one in Figure \ref{fig:stacked_gpu} with the timing breakdown for CPU compression (compression overhead + communication time + decompression overhead) as a function of node counts $N$.  At low-to-moderate
node counts ($N \leq 36$), the data volume per node is large
and CPU compression time exceeds by one order of magnitude the communication savings from reduced
message size, making the approach counter-productive.  At large node
counts ($N = 121$), message sizes are small enough that CPU
compression becomes feasible in principle, but large-scale NWP is a
weak-scaling workload: as node counts increase, the problem size
typically increases proportionally, so the data volume per node remains
approximately constant.  This means that CPU compression time does not
decrease as $N$ grows in a realistic production scenario, confirming
that GPU compression is the only viable pathway.
\subsection{Accuracy--Performance Trade-off}
\label{sec:results_tradeoff}
\begin{table}[hb]
  \centering
  \caption{Summary of the accuracy--performance trade-off at
    $N = 121$ nodes.  Mean relative error is reported
    over all $4^3$ ZFP blocks of the temperature field.
    Speedup is relative to the float32 baseline. Best values are underlined.}
  \label{table:tradeoff}
  \begin{tabular}{lcc}
    \toprule
    Strategy           & Speedup $\uparrow$       & Mean rel. error $\downarrow$     \\
    \midrule
    Float32 (baseline) & $1.0\times$              & ---                              \\
    Float16            & $2.05\times$             & $4 \times 10^{-4}$               \\
    ZFP rate-16        & $1.95\times$             & $\underline{2.5 \times 10^{-7}}$ \\
    ZFP rate-8         & $\underline{3.97\times}$ & $1 \times 10^{-4}$               \\
    \bottomrule
  \end{tabular}
\end{table}

Table \ref{table:tradeoff} summarizes the accuracy--performance
trade-off at $N = 121$ nodes, representative of a large-scale
operational run.  Float16 truncation achieves a $2.05\times$
speedup over float32 at a mean relative error of $4 \times 10^{-4}$.
ZFP rate-16 achieves roughly the same speedup ($1.95\times$) at a mean relative error of
$2.5 \times 10^{-7}$, a factor of $1600\times$ lower at a negligible additional cost.
ZFP rate-8 achieves a
$4.1\times$  speedup at a mean relative error of $10^{-4}$,
which is $4\times$ better than float16 at nearly double the speedup.
These results establish a clear improvement over float16 truncation. ZFP rate-16 achieves the same communication-time reduction at a negligible additional cost, while delivering a mean relative error that is three orders of magnitude lower.  ZFP rate-8 achieves an even better speedup at a mean relative error that is still better than float16 error.

\section{Future Directions}
\label{sec:future_directions}
In this work we only addressed the first
all-to-all transposition (by rows).  The second transposition (by columns) is
structurally analogous; the compression rationale and error analysis
apply equally.  End-to-end SHT performance would require both
transpositions to be compressed, and the combined speedup would be
somewhat smaller than the per-transposition figures reported here, as
local compute (FFT, Legendre transform) is not accelerated.
ZFP is a strong but general-purpose
compressor.  Other floating point compression libraries exist (see Section \ref{subsec:compression_scientific_data}), and custom-designed compressors tailored to atmospheric fields may achieve substantially better accuracy at
the same rate. Our results are therefore a conservative lower bound on
the achievable accuracy improvement.
We reported error at the communication
buffer level.  The effect of this error on the final spherical harmonic
coefficients and, downstream, on forecast accuracy over multi-day
integrations requires a full end-to-end experiment, see Kl\"ower et al. \cite{klower_2021__CompressingAtmosphericDataits} for a preliminary analysis of the effect of lossy compression on forecast accuracy.

\section{Conclusion}
\label{sec:conclusion}
We have presented a study on GPU-resident lossy
compression as an alternative to float16 truncation for reducing the
cost of MPI all-to-all pencil transpositions in the Spherical Harmonic
Transform.  Using measured GPU compression throughput from the cuZFP
backend and a SimGrid network simulation on a Dragonfly topology at
node counts representative of current operational NWP deployments, we
have shown that ZFP fixed-rate compression at 16 bits per value
achieves roughly the same communication-time reduction as float16 truncation
while delivering a mean relative error of $2.5 \times 10^{-7}$, a
factor of $1600\times$ lower than float16's $4 \times 10^{-4}$ on
the DYAMOND temperature field.  ZFP at 8 bits per value achieves
approximately $1.93\times$ the speedup of float16 at a mean
relative error of $10^{-4}$, which is $4\times$ better than float16 at nearly double the speedup.
We corroborated the trend of these simulated results with real measurements on the CPU partition of the CRESCO8 cluster, grounding the study in a currently-deployed, production-representative system.
Works like \cite{klower_2021__CompressingAtmosphericDataits,tintoprims_2024__EffectLossyCompressionnumerical} suggest that achievable compression ratio for atmospheric data may be higher than what we claimed, and so the accuracy--performance trade-off may be even better than what we report here. However, this has to be further investigated in future work balancing compression ratio, compression time and downstream forecast accuracy.
These results establish that the accuracy--performance trade-off curve
of ZFP fixed-rate strictly dominates that of float16 truncation.
We have also shown that CPU-based compression is not viable as a drop-in
replacement, confirming that GPU-resident compression is essential.

Operational pseudospectral climate and NWP codes run on CPUs today, and
GPU-resident compressed communication is, to our knowledge, a
currently-unexplored option for them.  As these codes migrate to GPU,
we argue this is an opportunity the climate/NWP community should weigh
alongside the engineering cost of the migration itself.
\section*{Acknowledgements}
This work was partially funded under the NRRP, Mission 4 Component 2 Investment 1.4, by the European Union -- NextGenerationEU (proj. no. CN 00000013, CUP no. E63C22000970007).  Views and opinions expressed are however those of the author(s) only and do not necessarily reflect those of the European Union or The European Research Executive Agency.  Neither the European Union nor the granting authority can be held responsible for them.
\bibliographystyle{IEEEtran}
\bibliography{library}

@inproceedings{ayala_2020_heFFTe_HeFFTeHighlyEfficientFFT,
  title = {{{heFFTe}}: {{Highly Efficient FFT}} for {{Exascale}}},
  shorttitle = {{{heFFTe}}},
  booktitle = {Computational {{Science}} -- {{ICCS}} 2020},
  author = {Ayala, Alan and Tomov, Stanimire and Haidar, Azzam and Dongarra, Jack},
  editor = {Krzhizhanovskaya, Valeria V. and Z{\'a}vodszky, G{\'a}bor and Lees, Michael H. and Dongarra, Jack J. and Sloot, Peter M. A. and Brissos, S{\'e}rgio and Teixeira, Jo{\~a}o},
  year = 2020,
  pages = {262--275},
  publisher = {Springer International Publishing},
  address = {Cham},
  doi = {10.1007/978-3-030-50371-0_19},
  isbn = {978-3-030-50371-0},
  language = {en}
}

@article{bucha_2026_CHarm_CHarmPythonLibraryspherical,
  title = {{{CHarm}}: {{C}}/{{Python}} Library for Spherical Harmonic Transforms in Planetary Geodesy},
  shorttitle = {{{CHarm}}},
  author = {Bucha, Bla{\v z}ej},
  year = 2026,
  month = mar,
  journal = {Earth Science Informatics},
  volume = {19},
  number = {3},
  pages = {29},
  issn = {1865-0473, 1865-0481},
  doi = {10.1007/s12145-026-02076-z},
  urldate = {2026-05-07},
  language = {en}
}

@article{casanova_2025_SimGrid_LoweringEntryBarriersdeveloping,
  title = {Lowering Entry Barriers to Developing Custom Simulators of Distributed Applications and Platforms with {{SimGrid}}},
  shorttitle = {{{SimGrid}}},
  author = {Casanova, Henri and Giersch, Arnaud and Legrand, Arnaud and Quinson, Martin and Suter, Fr{\'e}d{\'e}ric},
  year = 2025,
  month = mar,
  journal = {Parallel Computing},
  volume = {123},
  pages = {103125},
  issn = {01678191},
  doi = {10.1016/j.parco.2025.103125},
  urldate = {2026-05-07},
  language = {en}
}

@inproceedings{cayrols_2022_TruncatedheFFTe_LossyAlltoallExchangeaccelerating,
  title = {Lossy All-to-All Exchange for Accelerating Parallel 3-{{D FFTs}} on Hybrid Architectures with {{GPUs}}},
  shorttitle = {Truncated {{heFFTe}}},
  booktitle = {2022 {{IEEE International Conference}} on {{Cluster Computing}} ({{CLUSTER}})},
  author = {Cayrols, Sebastien and Li, Jiali and Bosilca, George and Tomov, Stanimire and Ayala, Alan and Dongarra, Jack},
  year = 2022,
  month = sep,
  pages = {152--160},
  publisher = {IEEE},
  address = {Heidelberg, Germany},
  doi = {10.1109/CLUSTER51413.2022.00029},
  urldate = {2026-05-07},
  copyright = {https://doi.org/10.15223/policy-029},
  isbn = {978-1-6654-9856-2},
  language = {en}
}

@book{coiffier_2011_FundamentalsofNWPBook_FundamentalsNumericalWeatherPrediction,
  title = {Fundamentals of {{Numerical Weather Prediction}}},
  shorttitle = {Fundamentals of {{NWP Book}}},
  author = {Coiffier, Jean},
  year = 2011,
  month = dec,
  edition = {1},
  publisher = {Cambridge University Press},
  doi = {10.1017/CBO9780511734458},
  urldate = {2026-05-26},
  copyright = {https://www.cambridge.org/core/terms},
  isbn = {978-1-107-00103-9 978-0-511-73445-8}
}

@article{desensi_2024_ExploringGPU-to-GPUCommunication_ExploringGPUtoGPUCommunication2024,
  title = {Exploring {{GPU-to-GPU Communication}}: 2024 {{International Conference}} for {{High Performance Computing}}, {{Networking}}, {{Storage}} and {{Analysis}}, {{SC}} 2024},
  shorttitle = {Exploring {{GPU-to-GPU Communication}}},
  author = {De Sensi, Daniele and Pichetti, Lorenzo and Vella, Flavio and De Matteis, Tiziano and Ren, Zebin and Fusco, Luigi and Turisini, Matteo and Cesarini, Daniele and Lust, Kurt and Trivedi, Animesh and Roweth, Duncan and Spiga, Filippo and Di Girolamo, Salvatore and Hoefler, Torsten},
  year = 2024,
  journal = {Sc24: International Conference for High Performance Computing, Networking, Storage and Analysis},
  series = {International {{Conference}} for {{High Performance Computing}}, {{Networking}}, {{Storage}} and {{Analysis}}, {{SC}}},
  pages = {1--15},
  publisher = {IEEE Computer Society},
  issn = {9798350352924},
  doi = {10.1109/SC41406.2024.00039},
  urldate = {2026-05-26}
}

@article{di_2025__SurveyErrorBoundedLossyCompression,
  title = {A {{Survey}} on {{Error-Bounded Lossy Compression}} for {{Scientific Datasets}}},
  author = {Di, Sheng and Liu, Jinyang and Zhao, Kai and Liang, Xin and Underwood, Robert and Zhang, Zhaorui and Shah, Milan and Huang, Yafan and Huang, Jiajun and Yu, Xiaodong and Ren, Congrong and Guo, Hanqi and Wilkins, Grant and Tao, Dingwen and Tian, Jiannan and Jin, Sian and Jian, Zizhe and Wang, Daoce and Rahman, Md Hasanur and Zhang, Boyuan and Song, Shihui and Calhoun, Jon and Li, Guanpeng and Yoshii, Kazutomo and Alharthi, Khalid and Cappello, Franck},
  year = 2025,
  month = jun,
  journal = {ACM Computing Surveys},
  volume = {57},
  number = {11},
  pages = {287:1--287:38},
  issn = {0360-0300},
  doi = {10.1145/3733104},
  urldate = {2026-05-27}
}

@article{diffenderfer_2019_ZFPanalysis_ErrorAnalysisZFPCompression,
  title = {Error {{Analysis}} of {{ZFP Compression}} for {{Floating-Point Data}}},
  shorttitle = {{{ZFP}} Analysis},
  author = {Diffenderfer, James and Fox, Alyson L. and Hittinger, Jeffrey A. and Sanders, Geoffrey and Lindstrom, Peter G.},
  year = 2019,
  month = jan,
  journal = {SIAM Journal on Scientific Computing},
  volume = {41},
  number = {3},
  pages = {A1867-A1898},
  issn = {1064-8275, 1095-7197},
  doi = {10.1137/18M1168832},
  urldate = {2026-05-07},
  language = {en}
}

@article{ecmwf_2024_IFSDocumentationCY49R1-PartVI_IFSDocumentationCY49R1Part,
  title = {{{IFS Documentation CY49R1}} - {{Part VI}}: {{Technical}} and {{Computational Procedures}}},
  shorttitle = {{{IFS Documentation CY49R1}} - {{Part VI}}},
  author = {{ECMWF}},
  year = 2024,
  publisher = {ECMWF},
  doi = {10.21957/D52E494677},
  urldate = {2026-05-07}
}

@misc{ecmwf_2026__ECMWFNewsletter186Winter,
  title = {{{ECMWF Newsletter}} 186 - {{Winter}} 2025/26},
  author = {{ECMWF}},
  year = 2026,
  publisher = {European Centre for Medium-Range Weather Forecasts},
  doi = {10.21957/SAGX-YF02},
  urldate = {2026-05-07},
  archiveprefix = {European Centre for Medium-Range Weather Forecasts},
  language = {en}
}

@article{gong_2023_MGARD_MGARDMultigridFrameworkhighperformance,
  title = {{{MGARD}}: {{A}} Multigrid Framework for High-Performance, Error-Controlled Data Compression and Refactoring},
  shorttitle = {{{MGARD}}},
  author = {Gong, Qian and Chen, Jieyang and Whitney, Ben and Liang, Xin and Reshniak, Viktor and Banerjee, Tania and Lee, Jaemoon and Rangarajan, Anand and Wan, Lipeng and Vidal, Nicolas and Liu, Qing and Gainaru, Ana and Podhorszki, Norbert and Archibald, Richard and Ranka, Sanjay and Klasky, Scott},
  year = 2023,
  month = dec,
  journal = {SoftwareX},
  volume = {24},
  pages = {101590},
  issn = {23527110},
  doi = {10.1016/j.softx.2023.101590},
  urldate = {2026-05-07}
}

@article{gorski_2005_HEALPix_HEALPixFrameworkHighResolutionDiscretization,
  title = {{{HEALPix}}: {{A Framework}} for {{High-Resolution Discretization}} and {{Fast Analysis}} of {{Data Distributed}} on the {{Sphere}}},
  shorttitle = {{{HEALPix}}},
  author = {G{\'o}rski, K. M. and Hivon, E. and Banday, A. J. and Wandelt, B. D. and Hansen, F. K. and Reinecke, M. and Bartelmann, M.},
  year = 2005,
  month = apr,
  journal = {The Astrophysical Journal},
  volume = {622},
  number = {2},
  pages = {759},
  issn = {0004-637X},
  doi = {10.1086/427976},
  urldate = {2026-05-26},
  language = {en}
}

@misc{huang_2023__CompressingMultidimensionalWeatherclimate,
  title = {Compressing Multidimensional Weather and Climate Data into Neural Networks},
  author = {Huang, Langwen and Hoefler, Torsten},
  year = 2023,
  month = apr,
  number = {arXiv:2210.12538},
  eprint = {2210.12538},
  primaryclass = {cs.LG},
  publisher = {arXiv},
  doi = {10.48550/arXiv.2210.12538},
  urldate = {2026-05-26},
  archiveprefix = {arXiv}
}

@misc{huang_2024__OptimizedErrorcontrolledMPICollective,
  title = {An {{Optimized Error-controlled MPI Collective Framework Integrated}} with {{Lossy Compression}}},
  author = {Huang, Jiajun and Di, Sheng and Yu, Xiaodong and Zhai, Yujia and Zhang, Zhaorui and Liu, Jinyang and Lu, Xiaoyi and Raffenetti, Ken and Zhou, Hui and Zhao, Kai and Chen, Zizhong and Cappello, Franck and Guo, Yanfei and Thakur, Rajeev},
  year = 2024,
  month = jan,
  number = {arXiv:2304.03890},
  eprint = {2304.03890},
  publisher = {arXiv},
  doi = {10.48550/arXiv.2304.03890},
  urldate = {2026-05-07},
  archiveprefix = {arXiv},
  language = {en}
}

@article{ishioka_2018_ISPACK_NewRecurrenceFormulaEfficient,
  title = {A {{New Recurrence Formula}} for {{Efficient Computation}} of {{Spherical Harmonic Transform}}},
  shorttitle = {{{ISPACK}}},
  author = {Ishioka, Keiichi},
  year = 2018,
  journal = {Journal of the Meteorological Society of Japan. Ser. II},
  volume = {96},
  number = {2},
  pages = {241--249},
  issn = {0026-1165, 2186-9057},
  doi = {10.2151/jmsj.2018-019},
  urldate = {2026-05-07},
  language = {en}
}

@inproceedings{klocke_2025_ICONat1kmGordonBell_ComputingFullEarthSystem,
  title = {Computing the {{Full Earth System}} at 1km {{Resolution}}},
  shorttitle = {{{ICON}} at 1 Km {{Gordon Bell}}},
  booktitle = {Proceedings of the {{International Conference}} for {{High Performance Computing}}, {{Networking}}, {{Storage}} and {{Analysis}}},
  author = {Klocke, Daniel and Frauen, Claudia and Engels, Jan Frederik and Alexeev, Dmitry and Redler, Ren{\'e} and Schnur, Reiner and Haak, Helmuth and Kornblueh, Luis and Br{\"u}ggemann, Nils and Chegini, Fatemeh and R{\"o}mmer, Manoel and Hoffmann, Lars and Griessbach, Sabine and Bode, Mathis and Coles, Jonathan and Gila, Miguel and Sawyer, William and Calotoiu, Alexandru and Budanaz, Yakup and Mazumder, Pratyai and Copik, Marcin and Weber, Benjamin and Herten, Andreas and Bockelmann, Hendryk and Hoefler, Torsten and Hohenegger, Cathy and Stevens, Bjorn},
  year = 2025,
  month = nov,
  series = {{{SC}} '25},
  pages = {125--136},
  publisher = {Association for Computing Machinery},
  address = {New York, NY, USA},
  doi = {10.1145/3712285.3771789},
  urldate = {2026-05-26},
  isbn = {979-8-4007-1466-5}
}

@article{klower_2021__CompressingAtmosphericDataits,
  title = {Compressing Atmospheric Data into Its Real Information Content},
  author = {Kl{\"o}wer, Milan and Razinger, Miha and Dominguez, Juan J. and D{\"u}ben, Peter D. and Palmer, Tim N.},
  year = 2021,
  month = nov,
  journal = {Nature Computational Science},
  volume = {1},
  number = {11},
  pages = {713--724},
  publisher = {Nature Publishing Group},
  issn = {2662-8457},
  doi = {10.1038/s43588-021-00156-2},
  urldate = {2026-05-27},
  copyright = {2021 The Author(s)},
  language = {en}
}

@inproceedings{li_2010_2decomp-fft_2decompFftaHighlyscalable,
  title = {2decomp \& Fft-a Highly Scalable 2d Decomposition Library and Fft Interface},
  shorttitle = {2decomp-Fft},
  booktitle = {Cray User Group 2010 Conference},
  author = {Li, Ning and Laizet, Sylvain},
  year = 2010,
  pages = {1--13}
}

@article{liang_2023_SZ3_SZ3ModularFrameworkComposing,
  title = {{{SZ3}}: {{A Modular Framework}} for {{Composing Prediction-Based Error-Bounded Lossy Compressors}}},
  shorttitle = {{{SZ3}}},
  author = {Liang, Xin and Zhao, Kai and Di, Sheng and Li, Sihuan and Underwood, Robert and Gok, Ali M. and Tian, Jiannan and Deng, Junjing and Calhoun, Jon C. and Tao, Dingwen and Chen, Zizhong and Cappello, Franck},
  year = 2023,
  month = apr,
  journal = {IEEE Transactions on Big Data},
  volume = {9},
  number = {2},
  pages = {485--498},
  issn = {2332-7790},
  doi = {10.1109/TBDATA.2022.3201176},
  urldate = {2026-05-07}
}

@article{lindstrom_2014_ZFP_FixedRateCompressedFloatingPointArrays,
  title = {Fixed-{{Rate Compressed Floating-Point Arrays}}},
  shorttitle = {{{ZFP}}},
  author = {Lindstrom, Peter},
  year = 2014,
  month = dec,
  journal = {IEEE Transactions on Visualization and Computer Graphics},
  volume = {20},
  number = {12},
  pages = {2674--2683},
  issn = {1077-2626, 1941-0506, 2160-9306},
  doi = {10.1109/TVCG.2014.2346458},
  urldate = {2026-05-07},
  copyright = {https://ieeexplore.ieee.org/Xplorehelp/downloads/license-information/IEEE.html}
}

@misc{ma_2026_UCCL-Zip_UCCLZipLosslessCompressionSupercharged,
  title = {{{UCCL-Zip}}: {{Lossless Compression Supercharged GPU Communication}}},
  shorttitle = {{{UCCL-Zip}}},
  author = {Ma, Shuang and Lao, Chon Lam and Xu, Zhiying and Wang, Zhuang and Mao, Ziming and Meng, Delong and Zhen, Jia and Wu, Jun and Stoica, Ion and Wang, Yida and Zhou, Yang},
  year = 2026,
  month = apr,
  number = {arXiv:2604.17172},
  eprint = {2604.17172},
  publisher = {arXiv},
  doi = {10.48550/arXiv.2604.17172},
  urldate = {2026-05-07},
  archiveprefix = {arXiv}
}

@article{pekurovsky_2012_P3DFFT_P3DFFTFrameworkParallelcomputations,
  title = {{{P3DFFT}}: A Framework for Parallel Computations of {{Fourier}} Transforms in Three Dimensions},
  shorttitle = {{{P3DFFT}}},
  author = {Pekurovsky, Dmitry},
  year = 2012,
  month = jan,
  journal = {SIAM Journal on Scientific Computing},
  volume = {34},
  number = {4},
  pages = {C192-C209},
  issn = {1064-8275, 1095-7197},
  doi = {10.1137/11082748X},
  urldate = {2026-05-21}
}

@article{reinecke_2013_Libsharp_LibsharpSphericalHarmonictransforms,
  title = {Libsharp -- Spherical Harmonic Transforms Revisited},
  shorttitle = {Libsharp},
  author = {Reinecke, M. and Seljebotn, D. S.},
  year = 2013,
  month = jun,
  journal = {Astronomy \& Astrophysics},
  volume = {554},
  pages = {A112},
  publisher = {EDP Sciences},
  issn = {0004-6361, 1432-0746},
  doi = {10.1051/0004-6361/201321494},
  urldate = {2026-05-07},
  copyright = {\copyright{} ESO, 2013},
  language = {en}
}

@article{reinecke_2020_DUCC_DUCCDistinctlyUsefulCode,
  title = {{{DUCC}}: {{Distinctly Useful Code Collection}}},
  shorttitle = {{{DUCC}}},
  author = {Reinecke, Martin},
  year = 2020,
  month = aug,
  journal = {Astrophysics Source Code Library},
  pages = {ascl:2008.023},
  urldate = {2026-05-07}
}

@article{rolfo_2023_2decomp-fft2_2DECOMPFFTLibraryUpdatenew,
  title = {The {{2DECOMP}}\&{{FFT}} Library: An Update with New {{CPU}}/{{GPU}} Capabilities},
  shorttitle = {2decomp-Fft 2},
  author = {Rolfo, Stefano and Flageul, C{\'e}dric and Bartholomew, Paul and Spiga, Filippo and Laizet, Sylvain},
  year = 2023,
  month = nov,
  journal = {Journal of Open Source Software},
  volume = {8},
  number = {91},
  pages = {5813},
  publisher = {Open Journals},
  doi = {10.21105/joss.05813},
  urldate = {2026-05-18}
}

@inproceedings{romero_2022_CuDecomp_DistributedmemorySimulationsTurbulentflows,
  title = {Distributed-Memory Simulations of Turbulent Flows on Modern {{GPU}} Systems Using an Adaptive Pencil Decomposition Library},
  shorttitle = {{{CuDecomp}}},
  booktitle = {Proceedings of the {{Platform}} for {{Advanced Scientific Computing Conference}}},
  author = {Romero, Joshua and Costa, Pedro and Fatica, Massimiliano},
  year = 2022,
  month = jul,
  series = {{{PASC}} '22},
  pages = {1--11},
  publisher = {Association for Computing Machinery},
  address = {New York, NY, USA},
  doi = {10.1145/3539781.3539797},
  urldate = {2026-05-07},
  isbn = {978-1-4503-9410-9}
}

@misc{schaeffer_2012_SHTns_EfficientSphericalHarmonicTransforms,
  title = {Efficient {{Spherical Harmonic Transforms}} Aimed at Pseudo-Spectral Numerical Simulations},
  shorttitle = {{{SHTns}}},
  author = {Schaeffer, Nathana{\"e}l},
  year = 2012,
  month = feb,
  journal = {arXiv.org},
  doi = {10.1002/ggge.20071},
  urldate = {2026-05-07},
  howpublished = {https://arxiv.org/abs/1202.6522v5},
  language = {en}
}

@article{stevens_2019_DYAMOND_DYAMONDDYnamicsAtmosphericgeneral,
  title = {{{DYAMOND}}: The {{DYnamics}} of the {{Atmospheric}} General Circulation {{Modeled On Non-hydrostatic Domains}}},
  shorttitle = {{{DYAMOND}}},
  author = {Stevens, Bjorn and Satoh, Masaki and Auger, Ludovic and Biercamp, Joachim and Bretherton, Christopher S. and Chen, Xi and D{\"u}ben, Peter and Judt, Falko and Khairoutdinov, Marat and Klocke, Daniel and Kodama, Chihiro and Kornblueh, Luis and Lin, Shian-Jiann and Neumann, Philipp and Putman, William M. and R{\"o}ber, Niklas and Shibuya, Ryosuke and Vanniere, Benoit and Vidale, Pier Luigi and Wedi, Nils and Zhou, Linjiong},
  year = 2019,
  month = sep,
  journal = {Progress in Earth and Planetary Science},
  volume = {6},
  number = {1},
  pages = {61},
  issn = {2197-4284},
  doi = {10.1186/s40645-019-0304-z},
  urldate = {2026-05-13},
  language = {en}
}

@inproceedings{tian_2020_cuSZ_CuSZEfficientGPUBasedErrorBounded,
  title = {{{cuSZ}}: {{An Efficient GPU-Based Error-Bounded Lossy Compression Framework}} for {{Scientific Data}}},
  shorttitle = {{{cuSZ}}},
  booktitle = {Proceedings of the {{ACM International Conference}} on {{Parallel Architectures}} and {{Compilation Techniques}}},
  author = {Tian, Jiannan and Di, Sheng and Zhao, Kai and Rivera, Cody and Fulp, Megan Hickman and Underwood, Robert and Jin, Sian and Liang, Xin and Calhoun, Jon and Tao, Dingwen and Cappello, Franck},
  year = 2020,
  month = sep,
  series = {{{PACT}} '20},
  pages = {3--15},
  publisher = {Association for Computing Machinery},
  address = {New York, NY, USA},
  doi = {10.1145/3410463.3414624},
  urldate = {2026-05-07},
  isbn = {978-1-4503-8075-1}
}

@article{tintoprims_2024__EffectLossyCompressionnumerical,
  title = {The Effect of Lossy Compression of Numerical Weather Prediction Data on Data Analysis: A Case Study Using Enstools-Compression 2023.11},
  author = {Tint{\'o} Prims, Oriol and Redl, Robert and Rautenhaus, Marc and Selz, Tobias and Matsunobu, Takumi and Modali, Kameswar Rao and Craig, George},
  year = 2024,
  month = dec,
  journal = {Geoscientific Model Development},
  volume = {17},
  number = {24},
  pages = {8909--8925},
  publisher = {Copernicus GmbH},
  issn = {1991-959X},
  doi = {10.5194/gmd-17-8909-2024},
  urldate = {2026-05-26},
  language = {English}
}

@misc{unat_2024__LandscapeGPUCentricCommunication,
  title = {The {{Landscape}} of {{GPU-Centric Communication}}},
  author = {Unat, Didem and Turimbetov, Ilyas and Issa, Mohammed Kefah Taha and Sa{\u g}bili, Do{\u g}an and Vella, Flavio and De Sensi, Daniele and Ismayilov, Ismayil},
  year = 2024,
  publisher = {arXiv},
  doi = {10.48550/ARXIV.2409.09874},
  urldate = {2026-05-26},
  copyright = {Creative Commons Attribution 4.0 International},
  language = {en}
}

@article{vana_2017_SinglePrecisioninWeatherForecastingModels_SinglePrecisionWeatherForecasting,
  title = {Single {{Precision}} in {{Weather Forecasting Models}}: {{An Evaluation}} with the {{IFS}}},
  shorttitle = {Single {{Precision}} in {{Weather Forecasting Models}}},
  author = {V{\'a}{\v n}a, Filip and D{\"u}ben, Peter and Lang, Simon and Palmer, Tim and Leutbecher, Martin and Salmond, Deborah and Carver, Glenn},
  year = 2017,
  month = feb,
  journal = {Monthly Weather Review},
  volume = {145},
  number = {2},
  pages = {495--502},
  issn = {0027-0644, 1520-0493},
  doi = {10.1175/MWR-D-16-0228.1},
  urldate = {2026-05-26},
  language = {en}
}

@article{zhao_2023_MFFT_MFFTGPUAcceleratedHighly,
  title = {{{MFFT}}: {{A GPU Accelerated Highly Efficient Mixed-Precision Large-Scale FFT Framework}}},
  shorttitle = {{{MFFT}}},
  author = {Zhao, Yuwen and Liu, Fangfang and Ma, Wenjing and Li, Huiyuan and Peng, Yuanchi and Wang, Cui},
  year = 2023,
  month = sep,
  journal = {ACM Transactions on Architecture and Code Optimization},
  volume = {20},
  number = {3},
  pages = {1--23},
  issn = {1544-3566, 1544-3973},
  doi = {10.1145/3605148},
  urldate = {2026-05-07},
  language = {en}
}

@inproceedings{zhou_2021__DesigningHighPerformanceMPILibraries,
  title = {Designing {{High-Performance MPI Libraries}} with {{On-the-fly Compression}} for {{Modern GPU Clusters}}},
  booktitle = {2021 {{IEEE International Parallel}} and {{Distributed Processing Symposium}} ({{IPDPS}})},
  author = {Zhou, Q. and Chu, C. and Kumar, N. S. and Kousha, P. and Ghazimirsaeed, S. M. and Subramoni, H. and Panda, D. K.},
  year = 2021,
  month = may,
  pages = {444--453},
  publisher = {IEEE},
  address = {Portland, OR, USA},
  doi = {10.1109/IPDPS49936.2021.00053},
  urldate = {2026-05-07},
  copyright = {https://ieeexplore.ieee.org/Xplorehelp/downloads/license-information/IEEE.html},
  isbn = {978-1-6654-4066-0},
  language = {en}
}

@incollection{zhou_2022__AcceleratingMPIAlltoAllCommunication,
  title = {Accelerating {{MPI All-to-All Communication}} with {{Online Compression}} on {{Modern GPU Clusters}}},
  booktitle = {High {{Performance Computing}}},
  author = {Zhou, Qinghua and Kousha, Pouya and Anthony, Quentin and Shafie Khorassani, Kawthar and Shafi, Aamir and Subramoni, Hari and Panda, Dhabaleswar K.},
  editor = {Varbanescu, Ana-Lucia and Bhatele, Abhinav and Luszczek, Piotr and Marc, Baboulin},
  year = 2022,
  volume = {13289},
  pages = {3--25},
  publisher = {Springer International Publishing},
  address = {Cham},
  doi = {10.1007/978-3-031-07312-0_1},
  urldate = {2026-05-07},
  isbn = {978-3-031-07311-3 978-3-031-07312-0},
  language = {en}
}

\end{document}